\documentstyle[11pt,aaspp]{article}
\input{psfig}
\eqsecnum
\def\beq{\begin{equation}}
\def\eeq{\end{equation}}
\def\ref{\reference}
\def\simge{\mathrel{%
   \rlap{\raise 0.511ex \hbox{$>$}}{\lower 0.511ex \hbox{$\sim$}}}}
\def\simle{\mathrel{
   \rlap{\raise 0.511ex \hbox{$<$}}{\lower 0.511ex \hbox{$\sim$}}}}

\def\ts{\times}

\def\.{\mathaccent 95}
\def\a{\alpha}

\def\ga{\gamma}

\def\ep{\epsilon}

\def\Ga{\Gamma}
\def\De{\Delta}

\def\Om{\Omega}

\def\frac#1#2{{\textstyle{{#1}\over {#2}}}}

\def\lsim{\mathrel{\rlap{\lower4pt\hbox{\hskip1pt$\sim$}}
     \raise1pt\hbox{$<$}}}
\def\gsim{\mathrel{\rlap{\lower4pt\hbox{\hskip1pt$\sim$}}
    \raise1pt\hbox{$>$}}}
\def\sqr#1#2{{\vcenter{\vbox{\hrule height.#2pt
         \hbox{\vrule width.#2pt height#1pt \kern#1pt
         \vrule width.#2pt}
         \hrule height.#2pt}}}}

\newbox\grsign \setbox\grsign=\hbox{$>$} \newdimen\grdimen \grdimen=\ht\grsign
\newbox\simlessbox \newbox\simgreatbox
\setbox\simgreatbox=\hbox{\raise.5ex\hbox{$>$}\llap
     {\lower.5ex\hbox{$\sim$}}}\ht1=\grdimen\dp1=0pt
\setbox\simlessbox=\hbox{\raise.5ex\hbox{$<$}\llap
     {\lower.5ex\hbox{$\sim$}}}\ht2=\grdimen\dp2=0pt

\def\ref#1  {\noindent \hangindent=24.0pt \hangafter=1 {#1} \par}
\def\doublespace {\smallskipamount=6pt plus2pt minus2pt
                  \medskipamount=12pt plus4pt minus4pt
                  \bigskipamount=24pt plus8pt minus8pt
                  \normalbaselineskip=24pt plus0pt minus0pt
                  \normallineskip=2pt
                  \normallineskiplimit=0pt
                  \jot=6pt
                  {\def\smallskip {\vskip\smallskipamount}}
                  {\def\medskip   {\vskip\medskipamount}}
                  {\def\bigskip   {\vskip\bigskipamount}}
                  {\setbox\strutbox=\hbox{\vrule 
                    height17.0pt depth7.0pt width 0pt}}
                  \parskip 12.0pt
                  \normalbaselines}
\def\tm{\times}
\def\ref{\noindent}
\begin{document}

\title{On Fueling Gamma-Ray Bursts and Their Afterglows with Pulsars}

\author{Eric G. Blackman$^1$ and Insu Yi$^2$}

\affil{$^1$Institute of Astronomy, Madingley Road, Cambridge CB3 OHA, England;
blackman@ast.cam.ac.uk}
\affil{$^2$Institute for Advanced Study, Olden Lane, Princeton, NJ 08540; 
yi@sns.ias.edu}
\medskip
\centerline{(submitted to ApJL)}

\vskip 0.3cm
\begin{abstract}

Cosmological gamma-ray bursts (GRBs) and their afterglows seem to result from 
dissipation of bulk energy in relativistic outflows, but their engine has 
not been unambiguously identified.  The engine could be a young pulsar formed 
from accretion induced collapse with a dynamo amplified field. Elsewhere, we 
suggest that such a ``Usov type'' strong field pulsar may help  explain the 
bimodal distribution in GRB durations.  Here we discuss possible roles of a 
pulsar for the afterglow.  We derive the expected bolometric luminosity
decay.  The extracted rotational energy could dissipate by shocks or by large 
amplitude electromagnetic waves (LAEMW).  The simplest LAEMW approach 
predicts a slower decay in observed afterglow peak frequency and faster decay 
in flux than the simplest blast-wave model, though more 
complicated models of both can provide different dependences. LAEMW do not 
require the rapid magnetic field amplification demanded of the blast-wave 
approach because the emission originates from a nearly fixed radius.  
Different time dependent behavior of GRB and post-GRB emission is also 
predicted.  Observational evidence for a pulsar in a GRB would make some 
GRB engine models, such as neutron star mergers and black holes unlikely. 
Therefore, the question of whether a pulsar is present is an important 
one even if it could drive a canonical fireball.

\end{abstract}


\section{Introduction}

Gamma-ray bursts (GRB) and their afterglows are likely the result of energy 
dissipation from a relativistically expanding outflow (e.g. Rees 1997a).  
Recent observations of GRB and their afterglows in optical and radio 
require a total GRB energy output $\sim 10^{52}\Delta\Phi$ erg 
where $\Delta\Phi$ is the solid angle fraction for the opening of the outflow 
from which emission occurs (e.g. Waxman et al. 1997, Dar 1997, and references 
therein).  The observation of an intervening absorber at redshift of 0.835 
(Metzger el. al 1997) in GRB 970508, suggests that this burst, if not all, 
are cosmological.  Although cosmological fireball+blast-wave models   
provide somewhat successful fits to the GRB and afterglow emission
(e.g. Wijers et al. 1997, Waxman 1997, Katz \& Piran 1997), the engine
driving the outflow 
is unknown. This gives GRB a unique place in modern astronomy as
the only known source whose engine does not yet point to a 
convincing paradigm. Even the energy budget is not fully 
constrained since the extent of beaming is unknown (e.g. Rhoads 1997).  

%The fireball and its afterglow may to some extent 
%represent a generic description of how
%the GRB engine energy is dissipated, but there are a number
%of possibilities for what actually powers the engine.

Magnetic fields are essential to GRB emission and afterglows in most 
viable models.  Two paradigms for the GRB and/or  the afterglow can be 
distinguished (e.g. Katz \& Piran 1997). In the first, emission 
results from dissipation in an out-flowing fireball and magnetized blast-wave 
interactacting with the external medium, ``external shock model.'' In the 
second,  energy is transferred into particles well internal to the interface 
between the outflow and the ambient medium. Internal shock models are an 
example of the latter but the class is better described just 
by ``internal'' since  shocks are not necessarily required. It appears that 
external models are not at work for the actual GRB, but may be fundamental for 
the afterglow. Whether internal processes may account for 
both is unknown. One example of a ``non-shock'' internal 
process is large amplitude electromagnetic waves (LAEMW), which can  
extract rotational energy (e.g. Michel 1984, Usov 
1994) from strongly magnetized, millisecond (or sub-millisecond) pulsars (MSP)
and can be applied to GRB.

Several pulsar or compact object models (e.g. Usov 1992, 
Thompson 1994, Blackman et. al 1996, Kluzniak \& Ruderman 1997; 
Meszaros \& Rees 1997, Yi \& Blackman 1998)
share the feature that power is first extracted in the form of Poynting flux
and is later converted to gamma-rays and lower frequency photons. Since the 
Poynting flux energizes particles at large distances from the magnetized 
rotator, the required low baryon fraction (e.g. Fenimore et al. 1993) is 
maintained, and the gamma-ray transparency condition on the bulk Lorentz factor 
of the relativistic outflow ($\gamma>100$) is ensured. The Poynting flux 
extraction of Meszaros \& Rees (1997) comes from a rotating black hole+torus 
engine, so the extraction is transient, lasting until the torus falls into 
the hole (tens of seconds), not throughout the afterglow, like a pulsar.

Pulsars formed from accretion induced collapse (AIC), with a dynamo amplified 
field, are appealing for GRB because two classes of such bursts can result, 
depending on whether or not the initial pulsar forms with spin
above or below the gravitationally unstable limit (Usov 1992). 
We have suggested elsewhere (Blackman et al. 1996, Yi \& Blackman 1998,
cf. Katz \& Canel 1996) that these two classes might account for the bimodal 
duration distribution (e.g. Kouveliotou et al. 1993).
%Regardless of model details, the central engines could be  
%expected to possess rapid rotation with spin frequencies $\sim 10^4 s^{-1}$ 
%and strong magnetic fields $\sim 10^{15}G$ (Usov 1992, Meszaros \& Rees 1997).
%is a particularly appealing paradigm for the engine for several reasons: 
The pulsar also provides an alternative to explain the GRB afterglows.
Also, pulsar LAEMW emission mechanisms exist which 
do not require the amplification of magnetic field required in 
blast-wave shock models.

We emphasize that even though the presence of a pulsar may be fundamental,
the actual mechanism by which the pulsar ultimately transfers its energy into 
radiation is not well understood. If a blast-wave is produced, the result may 
be very similar to transient rotator models (e.g. Meszaros \& Rees 1997)
formed from neutron star mergers (e.g. Narayan et al. 1992). But if 
observational evidence for a pulsar is found, GRB would probably {\it not}
be the result of neutron star mergers since the compact object formed is 
likely too large to be a neutron star (Ruffert et al. 1997). The question 
of a pulsar presence is thus extremely important.

In this paper we first summarize an AIC (cf. Yi \& Blackman
1998; also see Dar et al. 1992, Thompson 1994) pulsar model. 
We then discuss how the electromagnetic energy might dissipate.
We calculate features of an LAEMW afterglow and compare to the external 
shock fireball. We then conclude, suggesting features that 
could help indicate a pulsar presence.

\section{Pulsar Model and Bolometric Luminosity Evolution}

When a white dwarf reaches the critical Chandrasekhar mass $\sim 1.4M_{\odot}$
through mass accretion, it collapses to a neutron star.
The probable mass accretion rate leading to AIC is ${\dot M}\simge 3\times
10^{18} g/s$ (e.g. Livio \& Truran 1992), 
and  lasts for $>10^6 yr$. (Such constraints 
disfavor binary systems where the secondary is also a white dwarf.) 
%The white dwarf magnetic fields interact with
%accretion flows and directly affect its subsequent spin evolution. 
%Given the long accretion time scales, the initial
%spin has little effect on the final 
%pre-collapse dwarf (Yi \& Blackman 1997).
For a pre-AIC white dwarf with  moment of inertia
$I_{wd}=10^{51} g cm^2$ and  radius $R_{wd}=10^9 cm$, the magnetic
field $B_{wd}$ and spin frequency $\Omega_{wd}$
are related to the post-AIC pulsar spin frequency and magnetic field
by $\Omega_{wd}=\Omega_{*}(I_{*}/I_{wd})$ and
$B_{wd}=B_{*}(R_{*}/R_{wd})^2$ where $I_{*}=10^{45} g cm^2$
and $R_{*}=10^6 cm$ are the moment of inertia and the radius of the
MSP. In order to create a MSP ($\Omega_{*}\sim 10^4 s^{-1}$) with
$B_*\sim 10^{15} G$ by flux-freezing, 
the pre-collapse white dwarf must have
$\Omega_{wd}\sim 10^{-2} s$ and $B_{wd}\sim 10^9 G$. Such strong
white dwarf fields have not been observed. 
Moreover, Yi \& Blackman (1998) 
have shown that such AIC inducing magnetized accretion is not compatible with 
such a white dwarf.
%due to efficient magnetic braking and spin-down
%during pre-AIC magnetized accretion phase. 
They find that the most likely AIC-produced MSP parameters satisfy
$ \Omega_{*}\sim 10^4 B_{*,11}^{-4/5} s^{-1} $
where $B_{*,11}=B_{*}/10^{11}G$. 

But field amplication is natural in the early phases
of the neutron star, either linearly by differential rotation
(Kluzniak \& Ruderman 1997)
or exponentially by dynamo amplification (Duncan \& Thopmson 1992).
The dynamo would undoubtedly be the more efficient mode if operating.  
In the Kluzniak \& Ruderman (1997) approach 
(which could in principle be applied to any magnetized
rotator) the growth of magnetic field from differential rotation 
leads to intermittent expulsion of field from the pulsar, possibly 
accounting for the spiky light curves of GRB.  
This requires linear magnetic field growth rather than 
(dynamo) exponential, otherwise the expulsions would 
be too rapid.

Duncan \& Thompson (1997) suggested that a young hot MSP is a favorable 
site for an $\alpha\omega$ dynamo due to vigorous convection
driven by a large neutrino flux. 
%$\nu\sim 1-100 cm^2 s^{-1}$.
The dynamo efficiency can be estimated by the 
Rossby number $ N_{R}=P_{*}/\tau_{conv}$, 
where $P_{*}=2\pi/\Omega_{*}$ is the MSP spin period
and $\tau_{conv}$ is the convective overturn time scale at the base of
the convection zone. If $N_R\simle 1$ in a 
turbulent medium, an efficient dynamo can result.
Duncan and Thompson (1992) give 
the convective overturn time as $\tau_{conv}\sim 
10^{-3} F_{39}^{-1/3} s$ where $F_{39}$ is the convective neutrino
heat flux in units of $10^{39} {\rm erg/s/cm^2}$.
They show that a $\sim 10^{15}G$ large scale field can be produced. 

%The uncertainty associated with the required spin rate for an efficient dynamo
%action, $\Omega_{dynamo}\sim 2\pi \tau_{conv}^{-1}\simle 6\times 10^3 s^{-1}$ 
%is largely due to the uncertainty in the effectiveness of the dynamo when 
%$N_{R}\sim 1$ and the uncertainty in the convective overturn time scale.
%The increase of $N_R$ from $\sim 1$ by an order of magnitude seems to quench
%the build-up of the strong fields (Simon 1990). Nevertheless, 
%Given the possibility of some dynamo action 
%at $N_R\simge 1$, the dynamo-generated field may still exist for 
%$\Omega_*\simle \Omega_{dynamo}$.

The total luminosity emitted from the young pulsar has two important terms. 
An electromagnetic (EM) term and a gravitational radiation (GR) 
term. When the initial 
pulsar spin exceeds a critical spin $\Omega_{crit}$, the GR 
luminosity term can dominate (e.g. Usov 1992). 
For a young and highly viscous pulsar, the GR 
drain does not have time to change the star's moment of inertia 
(Yi \& Blackman 1998), so GR proceeds through spin-down, 
and the pulsar follows the track of an unstable Maclaurin spheroid (e.g.
Chandrasekhar 1969).  This means that the EM luminosity is 
depleted on the spin-down time scale, 
enabling the AIC-pulsar model to account for the bimodality of GRB durations
(Blackman et al, 1996; Yi \& Blackman 1998, however, cf. Katz \& Canel 1996).
If the young NS initially followed the Jacobi rather than
Maclaurin track, it would instead spin up, and would not
naturally lead to a bimodal distribution.  However, the GR from these two paths 
may be
measurably distinguishable (Lai \& Shapiro 1995).

%The secular instability driven by the gravitational wave emission
%and damped by the shear viscosity perturbs the axi-symmetric star
%with a non-axisymmetric perturbation of the form
%(e.g. Wagoner 1984)

%The gravitational radiation depletes energy and 
%angular momentum while $\delta R_*$ remains nearly constant. 
%Eventually, the gravitational instability growth 
%time scale gets longer as spin-down continues until $\tau_{gw}>\tau_{vis}$
%occurs for $l=m=2$. For $s\sim 1$, $I_{*}\sim constant$, $\delta R_*\sim R_*$
%and $\Omega_{*}\sim \Omega_{crit}$, the spin-down time scale
%$\tau_{down}\sim \tau_{gw}$.

Assuming $\Omega_{crit}>\Omega_{dynamo}$, 
where $\Omega_{dynamo}$ is the spin required for dynamo
action, there
are two types of pulsars which can form GRB:
%those
%with $\Omega_{*}\ge\Omega_{crit}$ or those with
%$\Omega_{*}\le\Omega_{crit}$ (cf. Blackman et al.
%1996). 
%If MSPs are formed from AIC of magnetized white dwarfs,
%the initia l MSP frequencies are determined by the pre-collapse white
%dwarf magnetic field strengths (Yi \& Blackman 1998).
[1] Supercritical strong field rotator (SPS) 
with $\Omega_{*}>\Omega_{crit}>\Omega_{dynamo}$, and 
[2] Subcritical strong field rotators (SBS) 
$\Omega_{crit}>\Omega_{*}>\Omega_{dynamo}$.  
Since these classes have similar rotational frequencies, they
likely originate from similar pre-collapse conditions  
so the numbers of the two classes should be similar.
The EM luminosity available for particles
(quadrupole may also be important) is
\beq
L_{obs}=f L_{em}\sim 10^{51}(f/0.1)(\De\Phi/0.01)^{-1}(\Omega_*/10^4~s^{-1})^4
(R_*/10^6cm)^6(B_*/10^{15}G)^2 ~{\rm erg/s}, 
\eeq
where $f$ is the efficiency.
The luminosity from GR is given by 
\beq
L_{gw}=32G\epsilon^2I_{*}^2\Omega_{*}^6/5c^5
=7\times 10^{55}(M_*/M_\odot)(R_*/10^6 cm)^4(\Omega_*/10^4~s^{-1})^6\ep^2~
{\rm erg/s},
\eeq 
where $\epsilon$ is the eccentricity.
%\beq
%L_{em}\sim 2\times 10^{50}R_{*,6}^6B_{*,15}^2\Omega_{*,4}^4~ erg/s
%\eeq
%We expect that SPS's $L_{em}$ ($\propto \Omega_*^4$)
%is slightly larger than SBS's due to slightly larger $\Omega_*$ for SPS.
%This is apparently consistent with the observed luminosity difference 
%between the long bursts and short bursts (e.g. Fishman \& Meegan 1995).
Both SPS and SBS emit EM radiation. 
However for SPS, the 
spin-down, and hence the luminosity e-folding decrease, occurs on a time scale
$\tau_{gw}\simle 1s$ whereas in SBS, the spin-down and
luminosity decrease occur on a time scale $\tau_{em}\sim 10^2 s$.
SPS may be naturally related to the short GRB bursts and SBS to
long bursts of the observed (Kouveliotou et al. 1993) 
bimodal duration distribution.

Using (2-1) and (2-2), we can solve for the time evolution of the PSR spin 
for $L_{gw}$ and $L_{em}$ dominated regimes respectively.
%long it takes for this $\Omega$ to be reached.
Setting the kinetic energy loss $L_{kin}=(1/2)M_*R_*^2\Omega_*(d\Omega_*/dt)$
equal to $L_{gw}$, we use (2-2) to solve for
$\Omega_*$.  Separating variables and integrating gives
\beq
\Omega_*=10^4(\Omega_0/10^4~s^{-1})[1+2.8\ts 10^3(\Omega_0/10^4~s^{-1})^4
(M_*/M_\odot)(R/10^6 cm)^2\ep^2((t-t_0)/1s)]^{-1/4}~s^{-1},
\eeq
where $\Omega_0=\Omega_*(t=t_0)$ is the spin rate of the initially formed 
pulsar at time $t=t_0$.  
%For the parameters as scaled, setting 
%$\Omega_*=\Omega_{eq}$ gives $t_{eq}-t_0\sim t_{eq}=10^6sec$.
At $t=t_{eq}$, the spin evolution for SPS bursts changes
from being GR dominated to EM radiation dominated.  This time is likely of order 
$\tau_{gw}$ since
$\ep$ decreases rapidly after that, and $L_{gw}$
becomes unimportant.  
For the SBS, the spin down is always dominated by $L_{em}$.  

For the SBS bursts and for SPS bursts after $t_{eq}$, we  
set (2-1) equal to $L_{kin}$ and integrate, so 
\beq
\Omega_*=10^4(\Omega_0/10^4~s^{-1})[1+2\ts 10^{-2}(\Omega_0/10^4~s^{-1})^2
(R/10^6 cm)^4(B_*/10^{15}G)^2((t-t_{0,em})/1s)]^{-1/2}~s^{-1}.
\eeq
where $t_{0,em}$ equals $t_0$ for the SBS or $t_{eq}$ for the SPS.

By plugging (2-3) and (2-4) back into (2-1)  we
can find the observed  bolometric 
luminosity time dependence for GRB and afterglow.
For the GR dominated phase
\beq
L_{obs}\sim 10^{51}(f/0.1)(\De\Phi/0.01)^{-1}(R_*/10^6cm)^6(B_*/10^{15}G)^2
(\Omega_0/10^4~s^{-1})^4
\qquad\qquad\qquad
\eeq
$$
\qquad\qquad
\ts [1+2.8\ts 10^3(\Omega_0/10^4~s^{-1})^4
(M_*/M_\odot)(R/10^6 cm)^2\ep^2((t-t_0)/1s)]^{-1}~{\rm erg/s}, 
$$
while for the EM dominated phase 
\beq
L_{obs} \sim 10^{51}(f/0.1)(\De\Phi/0.01)^{-1}(R_*/10^6cm)^6(B_*/10^{15}G)^2
(\Omega_{0,em}/10^4~s^{-1})^4
\qquad\qquad
\eeq
$$
\times [1+2\ts 10^{-2}(\Omega_{0,em}/10^4~s^{-1})^2
(R/10^6 cm)^4 (B_*/10^{15}G)^2((t-t_{0,em})/1s)]^{-2}~{\rm erg/s},
$$
where $\Omega_{0,em}$ is the spin at the beginning of EM 
domination at $t=t_{0,em}$.

Note that in the very early phase of the SBS bursts, 
the spin-down driven emission remains approximately constant and does not
begin to decay as a power law until the critical time
$t\sim 50$s from (2-6).  Whereas for the SPS bursts, from (2-5), 
a $t^{-1}$ decay occurs immediately, followed by a $t^{-2}$ decay
from the $L_{em}$ dominated phase of (2-6).  
Thus SBS are in qualitative accordance with the data analysis of 
Fenimore (1997) in the sense that during the GRB the burst decays
more slowly than in the subsequent afterglow.  This is just an example
of how the same internal process can account for different time evolutions
of the burst and afterglow.

\section{Peak Frequency and Associated Luminosity Evolution from LAEMW}

%**pulsar luminosity decay for synchrotron emission..

%*why the $\gamma^3$ dependence in frequency like curvature radii?

%**does the low end match synchrotron? $\nu^{1/3}$ 

%**not a;; pulsar remnants are plerions. in fact few are..
%why is this??

%**arons model employs particle energy dominated at the
%shock...see his review

%**why MHD is no good..

%**differences between LAEMW and blast-wave.

%-no protons in former

%-energy is not released all at once

%-radiation may occur before shock is formed?

%-size of region of interest

%**initial phase of characteristic freq as fun of time
%for GR bursts it decays in a millisecond.

%**spectrum

The magnetic field in the pulsar engine acts as a drive belt, extracting 
energy as Poynting flux to a large distance where it is then
converted into particles. (The magnetic field
plays a similar role in the Meszaros \& Rees (1997) model.)
The extracted energy  could subsequently
dissipate in large amplitude electromagnetic waves
or within a fireball+shock. How this energy is dissipated or converted 
into particles is not clear and may depend 
on the proton fraction of the outflow.
If there is a negligible proton fraction, then LAEMW  can propagate and 
accelerate pair plasma blobs (Asseo et al. 1978, Blackman et al. 1996).  
We discuss this and the associated afterglow below.

Avoiding the runaway pair production that would make a blob optically
thick to gamma rays, requires $\ga>100$ (e.g. Krolik \& Pier 1991).
Such pair plasma blobs moving along the magnetic dipole axis 
with $\ga> 10^4$ might be produced in PSR magnetospheres (Usov 1994) 
by the LAEMW.  The LAEMW propagate outside a radius, $r_{ff}$, where the density
drops below that which can sustain a Goldreich-Julian (Goldreich \& Julian
1969) charge density, and thus where flux-freezing and force-free conditions 
are broken. This gives (Usov 1994)
%$$r_{ff}\sim 10^{12}(B_{s0}/10^{13})^{1/2}(\Om_*/10^4{\rm 
%sec^{-1}})^{1/2},\eqno()$$
\beq
r_{ff}\sim 7.5\times 10^{13}(f/0.1)^{1/4}(\De\Phi/0.01)^{-1/4}
(B_{*}/10^{15} G)^{1/2}
(\Om_*/10^4{\rm s^{-1}})^{1/2}~{\rm cm},
\eeq
where we have included the beaming.
The associated pair plasma number density is
\beq
n_{ff}\sim 1.5\times 10^8(f/0.1)^{-3/4}(\De\Phi/0.01)^{3/4}
(R_*/10^6{\rm cm})(B_{*}/10^{15}{\rm G})^{1/2}(\Om_*/10^4{\rm 
s^{-1}})^{5/2}~{\rm cm^{-3}}.
%\eqno(15)
\eeq
Electron acceleration at  $r_{ff}$ is characterized by
$\sigma_{ff}$, defined by (Usov 1994; Michel 1984) 
\beq
\sigma_{ff}\equiv L_{dip}/(mc^2{\dot N}_{ff})\sim 5.3 \tm
10^7(f/0.1)^{-3/4}(\De\Phi/0.01)^{3/4}(R_*/10^6{\rm cm})^5
(B_{*}/10^{15}G)^{1/2}(\Om_*/10^4{\rm s^{-1}})^{1/2},
%\eqno(16)
\eeq
where ${\dot N}_{ff}=4\pi r_{ff}^2cn_{ff}$ is the electron flux 
and $m_e$ is the electron mass.
%pair plasma mass density, and $\phi\sim
%R_*^2B_{s0}(\Om_*R_*/c)$ is the radial magnetic flux.  
By solving the equations of motion for a particle in a pulsar wind zone
subject to electromagnetic forces,  it has been shown that 
relativistic electromagnetic 
waves of frequency $\Om_*$ can accelerate pair plasma to
(Michel 1984, Asseo et al. 1978)
$\ga \sim \sigma_{ff}^{2/3}\sim 10^6(R_*/10^6{\rm cm})^{10/3}
(B_{*}/10^{15}G)^{1/3}(\Om_*/10^4{\rm s^{-1}})^{1/3},$
and the resulting emission is beamed within solid angle $\sim
\ga^{-2}$ from the direction of wave propagation (Asseo et al. 1978).
The characteristic emitted frequency  of the synchro-Compton radiation 
is (Michel 1984) 
%is proportional to  $\ga^3$ or $\sigma_{ff}^2$ (like
%curvature radiation, eg. Beskin et al. 1993).  
\beq
\nu_{c} \sim ln(r_{ff}/R_*)\Omega_* \sigma_{ff}^2\sim 5\times 
10^{20}(f/0.1)^{-3/2}(\De\Phi/0.01)^{3/2}
(\Om_*/10^4{\rm s^{-1}})^2(R_*/10^6cm)^{10}(B_*/10^{15}G)~{\rm Hz}.
\eeq
%\eqno(18)
%with a tail to
%\beq
%\sigma_{ff}^{4/3}[eB_*(r_{ff})/(m_ec)]\sim 10^{24}{\rm sec^{-1}}
%(B_{s0}/10^{12}{\rm G})^{7/6}(R_*/10^6{\rm cm})^{23/3}
%(\Om_*/10^4{\rm sec^{-1}})^{1/6}.
%\eqno(19)
%\eeq
%frequency $\sim 10^{20}sec^{-1}(\Om_*/10^4sec^{-1})(\eta^{2/3}/(2{\rm
%x} 10^5)^3$.  

Using (3-4), (2-3) and (2-4), and assuming only $\Omega_*$ changes with time, 
we obtain different dependences of $\nu_c(t)$ 
for the GR and electromagnetic dominated phases. For the GR phase 
\beq
\nu_{c}=\nu_{c0}[1+2.8\times 10^3(\nu_{c0}/5\ts 10^{20} Hz)
(R_*/10^6cm)^{-18}(B_*/10^{15})^2(M/M_\odot)\ep^2((t-t_0)/1s)]^{-1/2},
\eeq
where $\nu_{c0}=\nu_c(\Omega_*=\Omega_0)$, while for the electromagnetic 
dominated regime
\beq
\nu_{c}=\nu_{c0,em}[1+2 \times 10^{-2}(\nu_{c0,em}/5\ts 10^{20} Hz)
(R_*/10^6cm)^{-9}(B_*/10^{15})((t-t_{0,em})/1s)]^{-1},
\eeq
where $\nu_{c0,em}=\nu_c(\Omega_*=\Omega_{0,em})$.
Note that when (3-5) applies, the total emission
is actually dominated by GR which peaks at 
10-100 Hz (Lai \& Shapiro 1995).

We can also compute the time decay of the luminosity (or flux)
at the peak electromagnetic emission frequency $\nu_c$.  
Assuming the form $L_\nu=L_{\nu_c}(\nu/\nu_c)^{-\a}$, 
with $\a>1$, we have $L_{obs}\sim L_{\nu_c}\nu_c$. 
Using $L_{\nu_c}\sim \nu_{c}^{-1}L_{obs}$, and
for the GR dominated phase using (3-4) and (3-5)
in (2-1) we get  
\beq
L_{\nu_c}
\sim 10^{51}(f/0.1)(\De\Phi/0.01)^{-1}(R_*/10^6cm)^6(B_*/10^{15}G)^2
(\Omega_0/10^4~s^{-1})^4
\qquad\qquad\qquad
\eeq
$$
\qquad\qquad
\times [1+2.8\ts 10^3(\Omega_0/10^4~s^{-1})^4
(M_*/M_\odot)(R/10^6 cm)^2\ep^2((t-t_0)/1s)]^{-1/2}~{\rm erg/s}.
$$
For the electromagnetic dominated phase using (3-4) and (3-6)
in (2-1) we get  
%$F_{\nu_c}\propto L_{\nu_c}\sim \nu_{c0,em}^{-1}$, 
\beq
L_{\nu_c}\sim10^{51}(f/0.1)(\De\Phi/0.01)^{-1}(R_*/10^6cm)^6(B_*/10^{15}G)^2 
(\Omega_{0,em}/10^4~s^{-1})^4\qquad\qquad\qquad
\eeq
$$
\qquad\qquad
\times [1+2\ts 10^{-2}(\Omega_{0,em}/10^4~s^{-1})^2
(R/10^6 cm)^4(B_*/10^{15}G)^2((t-t_{0,em})/1s)]^{-1}~{\rm erg/s}.
$$
%For the SPB, therefore, we would expect two distinct phases of time 
%dependence of the afterglow. That corresponding to the GR
%dominated phase, and the subsequent electromagnetic phase.
The evolution of the critical frequency and associated luminosity is less 
certain than the evolution of the bolometric luminosity when the latter is
directly related to pulsar spin-down.  Nevertheless we can still
speculate that the proton fraction in the outflow may be essential in 
determining what kind of dissipation occurs
and the LAEMW case discussed above is most relevant when the outflow
contains very few. Simulations of 
Hoshino et al. (1992) show that collisionless 
relativistic shocks produce a significant non-thermal 
lepton tail only when the protons comprise $>5$\% of the material.
In the proton enriched case, shock acceleration of electrons 
might be the most likely GRB
producing mechanism.  The simplest fireball evolution (e.g. Waxman 1997) 
leads to a $t^{-2}$ bolometric luminosity dependence, just like the LAEMW, 
but predicts $\nu_c\propto t^{-3/2}$ and $F_{\nu_c}\propto t^{-1/2}$.
This does seem to be consistent with several afterglows (Wijers et al 1997; Waxman 1997). 

The characteristic ``smoothness''
of the outflow Lorentz factor $\Ga$ may also be important in distinguishing
what kind of dissipation occurs.  For ejecta emitted at two different
times with two different $\Gamma$  satisfying  
$\Gamma_2 > \Gamma_1 >> 1$, 
internal GRB producing shocks could form at an observer frame distance of
$c|t_2-t_1|\Gamma_1\Gamma_2$ (Rees \& Meszaros 1994)
and may dissipate the energy before other internal mechanisms like 
LAEMW acceleration occur.

%\section {Comparison to Observations and Predictions}

%-luminosity decrease with time of bursts

%-characteristic frequency decrease with time of several bursts

%-predicted duality between bursts of the two different types:
%different dependences on luminosity, different dependences
%on 

\section {Discussion}

A generic afterglow may not clearly reveal the GRB engine, 
much like supernova remnant physics is  
disconnected from the physics of stellar collapse. 
A pulsar GRB engine could lead to a fireball or 
alternative acceleration mechanisms like LAEMW that
dissipate a large fraction of EM energy before shocks form at
a nearly fixed radius.  
One advantage of this LAEMW approach is that there is no need for 
the extremely rapid magnetic field amplification in the expanding 
outflow required of the fireball for both 
external and the internal shock models (e.g. Rees \& Meszaros 1994):
For LAEMW, the radius $r_{ff}$ and density at which the particle acceleration and radiation can occur remains nearly constant 
while the characteristic emission frequency decreases
from its sensitivity to pulsar spin down.
Alternatively, the pulsar could also provide some of the required
magnetic field, if its role were instead to generate a blast-wave.  

The simplest LAEMW picture predicts slower 
afterglow critical frequency decays than a blast-wave.  However, 
variations in the blast-wave models might lead to slower decays 
as it is not necessarily the case that most of the outflow 
energy is carried by the flow moving with the largest Lorentz factor 
(Rees \& Mezaros 1997).  To complicate things further, both blast-waves 
and LAEM could result from pulsars.  The proton fraction of the
outflow may be important in determining the primary mode of particle
acceleration and emission.  
In short, variations of energy dissipation mechanisms 
in strongly magnetized pulsar outflows could lead to a variety of
critical frequency and peak flux evolutions.  This 
makes it hard to distinguish between a 
pulsar engine vs. that of a transient Poynting flux extraction model.

However, whether a pulsar lies in  GRB engines is important
because the presence would be incompatible 
with the resulting compact object mass from neutron
star merger models and neutron star equations of state (Ruffert et al. 1997).
Evidence for pulsars in GRBs would threaten the merger paradigm.
Several possibilities that could hint at a pulsar are:
[1] The pulsar rotation itself may be a source of $\sim 10^{-3} s$. 
variability during the burst.
%In order for the stellar rotation to show up as variability,
%it is required that the magnetic dipole axis is misaligned with
%the rotational axis. For large misalignment angles, the amplitudes
%of variable fluxes are expected to be large.
For SBS (the longer) bursts, the characteristic 
time scale of variability after 50s or so from (2-6) 
goes as $\tau_v \propto \Omega_*^{-1}\propto t^{1/2}$.
This trend may be measurable in X-rays. 
For SPS (the shorter) bursts, the trend is 
$\tau_v \propto \Omega_*^{-1}\propto t^{1/4}$, immediately, and
then $\tau_v \propto \Omega_*^{-1}\propto t^{1/2}$ after $\tau_{gw}$.
[2] Future evidence in the radio band (Rees 1997b) 
like violent fluctuations or coherent emission (e.g. Melrose 1993). 
For GRB970508, the predicted bolometric 
luminosity from the $t^{-2}$ decay after $t\sim 50$days is about equal to 
that required of the radio afterglow luminosity for GRB970508 which is 
$L_{obs}\sim 10^{41} (\nu/5GHz)(F_{\nu}/10^{-26}{\rm erg/s/cm^2})
(d/5Gpc)^2$ erg/s (Waxman et al. 1997): Using (2-6) for SBS, we have 
$L_{obs} \sim 10^{41}$ erg/s at $t=50$ days.  
However, the decreasing scintillation amplitude 
suggests radio source expansion (Goodman 1997, Waxman et al. 1997),  
so here any pulsar's role could have been mainly to drive a 
blast-wave, rather than LAEMW.
[3] The SPS class of GRBs would be accompanied by GR
absent from the longer SBS class.
%Provided that young gravitationally radiating 
%AIC-MSP deplete their rotational energy through spin-down as described
%herein, 
The detection of a significant GR excess from short GRBs would give
support to an AIC pulsar engine model.  
The GR signature for various evolutionary 
tracks of secularly unstable
AIC pulsars are unique and possibly measurable (Lai \& Shapiro 1995).
[4] The SPS bursts should show a $t^{-1}$ decay in bolometric
luminosity followed by a $t^{-2}$ decay after the gamma-ray phase.
The SBS bursts should show almost no decay during the bursts
and then a $t^{-2}$ decay after the GRB phase ends. This follows 
from (3-7) and (3-8).
[5] Observations might reveal a faint residual pulsar or X-ray accreting 
source well after the optical afterglow fades.  An AXAF
view of an afterglow could be revealing.  

%Given the complex burst time profiles (Fishman \& Meegan 1995), 
%it is interesting that the
%shortest variability time scale is comparable to the shortest GRB duration.
%In the MSP scenario, this is naturally explained as the MSP spin-down
%time scale can be comparable to MSP spin period.  Note also 
%the fact that the GRB engine in this model would remain a stable
%pulsar long after the gamma-ray emission. 

Finally, the AIC rate for pulsar formation is unknown, but could be comparable to the local supernova rate, i.e. $10^{-2} yr^{-1}$ per galaxy, the
observed GRB rate of $\sim 10^{-6}-10^{-5} yr^{-1}$ per galaxy
can be amply explained by the AIC model with modest beaming.
These numbers are not inconsistent with the possibility that GRB
signature  {\it all} AIC events.

%if $\Delta\Omega\sim 10^{-5}-10^{-3}$
%which is largely consistent with the luminosity requirement if
%$\xi\sim 10^{-3}$.

E.B. thanks M. Rees and R. Wijers for discussions and 
I. Y. acknowledges support from SUAM Foundation,

\ref {} Asseo, E., Kennel, C.F. 1978, \& Pellat,R., A.\& A., 65, 401.

%\ref{} Balbinski, E. \& Schutz, B. F. 1982, MNRAS, 200, 43

\ref{} Blackman, E.G., Yi, I., \& Field, G. B. 1996, ApJ, 473, L79

%\ref{} Blair, D. 1989, in Gravitational Wave Data Analysis ed. B. F. Schutz
%(Dordrecht: Kluwer)

\ref{} Chandrasekhar, S. 1969, Ellipsoidal Figures of Equilibrium
(New Haven: Yale Univ. Press)

%\ref{} Comins, N. 1979, MNRAS, 189, 233

\ref{} Dar, A. 1997, preprint (astro-ph/9709231)

\ref{} Dar, A., Kozlovsky, B. Z., Nussinov, S., \& Ramaty, R. 1992, ApJ, 388, 
164

\ref{} Duncan, R. C. \& Thompson, C. 1992, ApJ, 392, L9

\ref{} Fenimore, E.E., Epstein, R. I., \& Ho, C. 1993, A\&AS, 97, 59

\ref{} Fenimore, E.E. 1997, submitted to ApJ.

%\ref{} Fishman, G. J. \& Meegan, C. A. 1995, ARA\&A, 33, 415

%\ref{} Friedman, J. L. 1983, PRL, 51, 11

\ref {} Goldreich, P. \& Julian, W. H. 1969, ApJ 157, 869.

\ref {} Goodman, J, 1997, New Astronomy, 2, 449.

\ref{} Hoshino, M. et al. 1992, ApJ, 390, 454.  
%,  J. Arons, Y. A. Gallant \& A. B. Langdon,

\ref{} Katz, J. I. \& Canel, L. M. 1996, ApJ, 471, 915

\ref{} Katz, J.I. \& Piran, T. 1997, astro-ph/9712242 

\ref{} Kluzniak, W. \& Ruderman, M. 1997, preprint (astro-ph/9712320)

\ref{} Kouveliotou C., et al. 1993, ApJ, 413, L101

\ref{} Krolik, J. H. \& Pier, E. A. 1991, ApJ, 373, 277

%\ref{} Lindblom, L. 1986, ApJ, 303, 146

\ref{} Lai, D. \& Shapiro, S.L. 1995, ApJ, 442, 559

\ref{} Livio, M. \& Truran, J. W. 1992, ApJ, 389, 695

%\ref{} Martin, P.G., \& Rees, M.J., MNRAS, 1979, 189, 19

\ref{} Melrose, D.B. 1993, in {\it Pulsars as Physics Labs}, eds. R.D. 
Blandford et.al (Oxford: New York)

\ref{} Meszaros, P. \& Rees, M. J. 1997, ApJ, 482, L29

\ref{} Metzger, M. et al. 1997, Nature, 387, 878

\ref{} Michel, F.C. 1984,  ApJ, 284, 384

\ref{} Narayan, R., Paczynski, B., \& Piran T. 1992, ApJ, 395, L83

%\ref{} Pazcynski, B. 1997, preprint (astro-ph/9710086)

\ref{} Rees, M. J. 1997a, preprint (astro-ph/9701162)

\ref{} Rees, M.J, 1997b, private communication

\ref{} Rees, M.J. \& Meszaros, P. 1994, ApJ, 430, L93

\ref{} Rees, M.J. \& Meszaros, P. 1997, preprint (astro-ph/9712252)

\ref{} Rhoads, J. 1997, ApJL, 487, L1

\ref{} Ruffert et. al 1997, A\& A, 319 122

%\ref{} Sari, R. \& Piran, T. 1995, ApJ, 455, L143

%\ref{} Simon, T. 1990, ApJ, 359, L11

\ref {} Thompson, C. 1994, MNRAS, 270, 480.

\ref{} Usov, V.V. 1992, Nature, 357, 452

\ref{} Usov, V.V. 1994, in  Gamma-Ray Bursts, AIP Conference
Proc. 307, (AIP: New York) p552.

%\ref{} Wagoner, R. V. 1984, ApJ, 278, 345

\ref{} Waxman, E. 1997, ApJ, 489, L33

\ref{} Waxman, E., Kulkarni, S.R., \& Frail, D.A. 1997, 
preprint (astro-ph/9709199).

\ref{} Wijers, R.A.M.J, Meszaros, P. \& Rees, M.J., 1997, MNRAS 288, L51.

\ref{} Yi, I. \& Blackman, E.G., 1998, ApJL, in press (astro-ph/9710149)

%\ref{} Yi, I. \& Blackman, E. G. 1997b, ApJ, 482, 383

%\end{references}
\clearpage

\end{document}